\documentstyle[prb,aps,epsf]{revtex}
\date{\today}
\title{Basis Functions for \\
Linear-Scaling 
First-Principles Calculations}
\author{E. Hern\'{a}ndez and M. J. Gillan
\\
Physics Department, Keele University, Staffordshire ST5 5BG, U.K. 
\medskip \\
C. M. Goringe \\
Materials Department, Oxford University, Oxford OX1 3PH, U.K.}
\begin{document}
\maketitle

\begin{abstract}
\begin{sloppypar}
In the framework of a recently reported linear-scaling method for
density-functional-pseudopotential calculations, we investigate the
use of localized basis functions for such work. We propose a basis set
in which each local orbital is represented in terms of an array of
``blip functions'' on the points of a grid. 
We analyze the relation between
blip-function basis sets and the plane-wave basis used in standard
pseudopotential methods, derive criteria for the approximate equivalence
of the two, and describe practical tests of these criteria. Techniques
are presented for using blip-function basis sets in linear-scaling
calculations, and numerical tests of these techniques are reported for 
Si crystal using both local and non-local pseudopotentials. We find
rapid convergence of the total energy to the values given by standard
plane-wave calculations as the radius of the linear-scaling localized 
orbitals is increased.
\end{sloppypar}
\end{abstract}

\twocolumn

\section{Introduction}
First-principles calculations based on density functional 
theory~\cite{hohenberg:kohn,kohn:sham} (DFT)
and the pseudo\-potential method are widely used for studying
the energetics, structure and dynamics of solids and 
liquids~\cite{gillan,galli:pasquarello,payne}. In the
standard approach, the occupied Kohn-Sham orbitals are expanded
in terms of plane waves, and the ground state is found by minimizing
the total energy with respect to the plane-wave 
coefficients~\cite{car:parrinello}.
Calculations on systems of over a hundred atoms with this approach
are now quite common. However, it has proved difficult to go to very
much larger systems, because the computational effort in this approach
depends on the number of atoms $N$ at least as $N^2$, and asymptotically
as $N^3$. Because of this limitation, there has been a vigorous effort
in the past few years to develop linear-scaling 
methods~\cite{yang1,yang2,yang3,yang4,yang5,baroni,galli,mauri1,mauri2,mauri3,ordejon1,ordejon2,li,nunes,hierse,hernandez:gillan,hernandez:gillan:goringe,ordejon:artacho:soler}
-- methods
in which the effort depends only linearly on the number of atoms. 
We have recently described a general theoretical framework for developing
linear-scaling self-consistent DFT 
schemes~\cite{hernandez:gillan,hernandez:gillan:goringe}. 
We presented one practical way of implementing
such a scheme, and investigated its performance for crystalline silicon.
Closely related ideas have been reported by other 
authors~\cite{hierse,ordejon:artacho:soler} -- an overview
of work on linear-scaling methods was given in the Introduction of
our previous paper~\cite{hernandez:gillan:goringe}.

The practical feasibility of linear-scaling DFT techniques is thus
well established. However, there are still technical problems to be solved
before the techniques can be routinely applied. Our aim here is to
study the problem of representing the localized orbitals that appear
in linear-scaling methods (support functions in our terminology) --
in other words, the problem of basis functions. 
To put this in context, we recall briefly the main
ideas of our linear-scaling DFT method.

Standard DFT can be expressed in terms of the Kohn-Sham density matrix
$\rho ( {\bf r}, {\bf r}^\prime )$. The total-energy functional
can be written in terms of $\rho$, and the ground state is obtained
by minimization with respect to $\rho$ subject to two constraints:
$\rho$ is idempotent (it is a projector, so that its eigenvalues are
0 or 1), and its trace is equal to half the number of electrons.
Linear-scaling behavior is obtained by imposing a limitation on the
spatial range of $\rho$:
\begin{equation}
\rho ( {\bf r}, {\bf r}^\prime ) = 0 \; , \; \; \; | {\bf r} -
{\bf r}^\prime | > R_c \; .
\end{equation}
By the variational principle, we then get an upper bound $E ( R_c )$
to the true ground-state energy $E_0$. Since the true ground-state
density matrix decays to zero as $| {\bf r} - {\bf r}^\prime |
\rightarrow \infty$, we expect that $E ( R_c \rightarrow \infty ) = 
E_0$. To make the scheme practicable, we introduced the further
condition that $\rho$ be separable:
\begin{equation}
\rho ( {\bf r}, {\bf r}^\prime ) = \sum_{\alpha \beta} 
\phi_\alpha ( {\bf r} ) \, K_{\alpha 
\beta} \, \phi_\beta ( {\bf r}^\prime ) \; ,
\end{equation}
where the number of support functions $\phi_\alpha ( {\bf r} )$ is finite.
The limitation on the spatial range of $\rho$ is imposed by requiring
that the $\phi_\alpha ( {\bf r} )$ are non-zero only in localized
regions (``support regions'') and that the spatial range of 
$K_{\alpha \beta}$ is limited. In our method, the support regions are
centered on the atoms and move with them.

We have shown~\cite{hernandez:gillan,hernandez:gillan:goringe} 
that the condition on the eigenvalues of $\rho$ can be
satisfied by the method of Li, Nunes and Vanderbilt~\cite{li}
 (LNV): instead
of directly varying $\rho$, we express it as:
\begin{equation}
\rho = 3 \sigma * \sigma - 2 \sigma * \sigma * \sigma \; ,
\end{equation}
where the asterisk indicates the continuum analog of matrix multiplication.
As shown by LNV, this representation of $\rho$ not only ensures
that its eigenvalues lie in the range $[ 0 , 1 ]$, but it drives
them towards the values 0 and 1. In our scheme, the auxiliary
matrix $\sigma ( {\bf r}, {\bf r}^\prime )$ has the same type of
separability as $\rho$:
\begin{equation}
\sigma ( {\bf r}, {\bf r}^\prime ) = \sum_{\alpha \beta}
\phi_\alpha ( {\bf r} ) \, L_{\alpha \beta} \,
\phi_\beta ( {\bf r}^\prime ) \; .
\end{equation}
This means that $K$ is given by the matrix equation:
\begin{equation}
K = 3 L S L - 2 L S L S L \; ,
\end{equation}
where $S_{\alpha \beta}$ is the overlap matrix of support functions:
\begin{equation}
S_{\alpha \beta} = \int \! \! \mbox{d} {\bf r} \, \phi_\alpha \phi_\beta \; .
\label{eq:overlap}
\end{equation}
We can therefore summarize the overall scheme as follows. The total
energy is expressed in terms of $\rho$, which depends on
the separable quantity $\sigma$. The ground-state energy is obtained
by minimization with respect to the support functions 
$\phi_\alpha ( {\bf r} )$ and the matrix elements $L_{\alpha \beta}$,
with the $\phi_\alpha$ confined to localized regions 
centered on the atoms, and the
$L_{\alpha \beta}$ subject to a spatial cut-off.

The $\phi_\alpha ( {\bf r} )$ must  be allowed to vary freely
in the minimization process, just like the Kohn-Sham orbitals
in conventional DFT, and we must consider how to represent them. As always,
there is a choice: we can represent them either by their values on
a grid~\cite{chelikowsky1,chelikowsky2,chelikowsky3,gygi1,gygi2,gygi3,seitsonen,hamann1,hamann2,bernholc}, 
or in terms of some set of basis functions. 
In our previous work~\cite{hernandez:gillan,hernandez:gillan:goringe},
we used a grid representation. This was satisfactory for 
discussing the feasibility of linear-scaling schemes, but seems to us to
suffer from significant drawbacks. Since the
support regions are centered on the ions in our method, this means
that when the ions move, the boundaries of the regions will 
cross the grid points. In any simple grid-based scheme, this will
cause troublesome discontinuities. In addition, the finite-difference
representation of the kinetic-energy operator in a grid
representation causes problems at the boundaries of the regions.
A further point is that in a purely grid-based method we are almost certainly
using more variables than are really necessary. 
These problems have led us to consider
basis-function methods.

We describe in this paper a practical basis-function scheme for
linear-scaling DFT, and we study its performance in numerical
calculations. The basis consists of an array of localized functions
-- we call them ``blip functions''.
There is an array of blip functions for each support region, and the
array moves with the region. The use of such arrays of localized
functions as a basis for quantum calculations is not 
new~\cite{cho:arias:joannopoulos:lam,chen:chang:hsue,modisette,wei:chou}.
However, to our knowledge it
has not been discussed before in the context of linear-scaling calculations.

The plan of the paper is as follows. In Sec.~2, we emphasize
the importance of considering the relation between blip-function
and plane-wave basis sets, and we use this relation to analyze
how the calculated ground-state energy will depend on the width
and spacing of the blip functions. 
We note some advantages of using
B-splines as blip functions, and we then present some practical
tests which illustrate the convergence of the ground-state energy
with respect to blip width and spacing. 
We then go on (Sec.~3) to discuss the
technical problems of using blip-function basis sets in linear-scaling
DFT. We report the results of
practical tests, which show explicitly how the ground-state
energy in linear-scaling DFT converges to the value
obtained in a standard plane-wave calculation. Section~4 gives
a discussion of the results, and presents our conclusions. Some 
mathematical derivations are given in an appendix.

\section{Blip functions and plane waves}
\subsection{General considerations}
Before we focus on linear-scaling problems, we need to set down some
elementary ideas about basis functions. To start with, we therefore
ignore the linear-scaling aspects, and we discuss the general problem
of solving Schr\"{o}dinger's equation using basis functions. It is
enough to discuss this in one dimension, and we assume a periodically
repeating system, so that the potential $V(x)$ acting on the
electrons is periodic: $V(x+t) = V(x)$, where $t$ is any
translation vector. Self-consistency questions are irrelevant at this
stage, so that $V(x)$ is given. The generalization to 
three-dimensional self-consistent calculations will be straightforward.

In a plane-wave basis, the wavefunctions $\psi_i (x)$ are expanded as
\begin{equation}
\psi_i (x) = L^{- 1/2} \sum_G c_{i G} \, 
\exp ( i G x ) \; ,
\end{equation}
where the reciprocal lattice vectors of the repeating geometry
are given by $G = 2 \pi n / L$ ($n$ is an integer), and we include
all $G$ up to some cut-off $G_{\rm max}$.
We obtain the ground-state energy
$E( G_{\rm max} )$ in the given basis by minimization with respect to
the $c_{i G}$, subject to the constraints of orthonormality.
For the usual variational reasons, $E ( G_{\rm max} )$ is a monotonically
decreasing function of $G_{\rm max}$ which tends to the exact value $E_0$ as
$G_{\rm max} \rightarrow \infty$.

Now instead of plane waves we want to use an array of spatially localized
basis functions (blip functions). 
Let $f_0 (x)$ be some localized function, and denote by
$f_\ell (x)$ the translated
function $f_0 (x - \ell a )$, where $\ell$ is an integer
and $a$ is a spacing, which is chosen so that $L$ is an exact multiple
of $a$: $L= M a$. We use the array of blip functions $f_\ell (x)$ as a basis.
Equivalently, we can use any independent linear combinations of the 
$f_\ell (x)$. In considering the relation between blip functions and
plane waves, it is particularly convenient to work with 
``blip waves'', $\chi_G (x)$, defined as:
\begin{equation}
\chi_G (x) = A_G \sum_{\ell = 0}^{M - 1} f_\ell (x) \exp ( i G R_\ell ) \; ,
\end{equation}
where $R_\ell = \ell a$, and $A_G$ is some normalization constant.

The relation between blip waves and plane waves can be analyzed
by considering the Fourier representation of $\chi_G (x)$. It is
straightforward to show that $\chi_G (x)$ has Fourier components only
at wavevectors $G + \Gamma$, where $\Gamma$ is a reciprocal
lattice vector of the blip grid: $\Gamma = 2 \pi m / a$ ($m$ is an integer).
In fact:
\begin{equation}
\chi_G (x) = ( A_G / a) \sum_{\Gamma} \hat{f} (G + \Gamma )
\exp \left( i ( G + \Gamma ) x \right) \; ,
\label{eq:blipwave}
\end{equation}
where $\hat{f} (q)$ is the Fourier transform of $f_0 (x)$:
\begin{equation}
\hat{f} (q) = \int_{- \infty}^{\infty} \mbox{d}x \, f_0 (x) \, e^{i q x} \; .
\label{eq:fourier}
\end{equation}

At this point, it is useful to note that for some
choices of $f_0 (x)$ the blip-function basis set is exactly equivalent
to a plane-wave basis set. For this to happen, $\hat{f} (q)$
must be exactly zero beyond some cut-off wavevector $q_{\rm cut}$.
Then provided $q_{\rm cut} \geq G_{\rm max}$ and provided
$q_{\rm cut} + G_{\rm max} < 2 \pi / a$, all the $\Gamma \neq 0$
terms in Eq.~(\ref{eq:blipwave}) will vanish and all 
blip waves for $- G_{\rm max} 
\leq G \leq G_{\rm max}$ will be identical to plane-waves. (Of course, we
must also require that $\hat{f} (q) \neq 0$ for $| q | \leq G_{\rm max}$.)

Our main aim in this Section is to determine how the total energy
converges to the exact value as the width and spacing of the
blip functions are varied. The spacing is controlled by varying $a$,
and the width is controlled by scaling each blip function: $f_0 (x)
\rightarrow f_0 (sx)$, where $s$ is a scaling factor. In the case
of blip functions for which $\hat{f} (q)$ cuts off in the way just
described, the convergence of the total energy is easy to describe.
Suppose we take a fixed blip width, and hence a fixed wavevector
cut-off $q_{\rm cut}$. If the blip spacing $a$ is small enough so that
$q_{\rm cut} < \pi / a$, then it follows from what we have said
that the blip basis set is exactly equivalent to a plane-wave basis set
having $G_{\rm max} = q_{\rm cut}$. This means that the total
energy is equal to $E( q_{\rm cut} )$ and is completely independent
of $a$ when the latter falls below the threshold value
$a_{\rm th} = \pi / q_{\rm cut}$.  This is connected with the fact
that the blip basis set becomes over-complete when $a < a_{\rm th}$: there
are linear dependences between the $M$ blip functions 
$f_\ell (x)$.

It follows from this that the behavior of the total energy as a function
of blip spacing and blip width is as shown schematically in Fig.~1.
As the width is reduced, the cut-off $q_{\rm cut}$ increases in 
proportion to the scaling factor $s$, so that the threshold spacing
$a_{\rm th}$ is proportional to the width. The energy 
value $E( q_{\rm cut} )$ obtained for $a < a_{\rm th}$ decreases 
monotonically with the width, as follows from the monotonic decrease
of $E( G_{\rm max} )$ with $G_{\rm max}$ for a plane-wave basis set.
Note that in Fig.~1 we have shown $E$ at fixed width as decreasing
monotonically with $a$ for $a > a_{\rm th}$. In fact, this may not always
happen. Decrease of $a$ does not correspond simply to addition of
basis functions and hence to increase of variational freedom: it also involves
relocation of the basis functions. However, what is true is that
$E$ for $a > a_{\rm th}$ is always greater than $E$ for $a < a_{\rm th}$,
as can be proved from the over-completeness of the blip basis set
for $a < a_{\rm th}$. At large spacings, the large-width blip basis is
expected to give the lower energy, since in this region the poorer 
representation of long waves should be the dominant source of error.

\begin{figure}[tbh]
\begin{center}
\leavevmode
\epsfxsize=8cm
\epsfbox{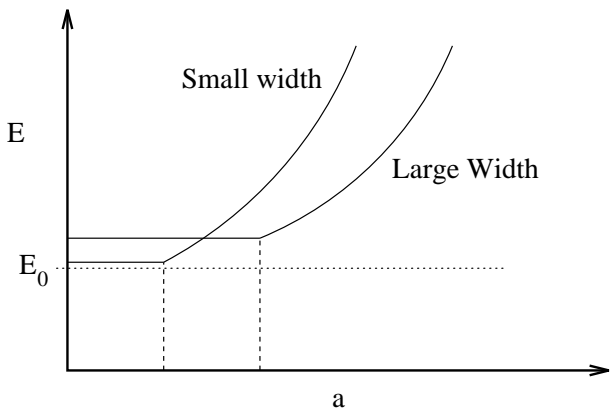}
\end{center}
\caption{Expected schematic form for the total ground-state energy
as a function of the blip-grid spacing for two different
blip widths, in the case where the Fourier components of the
blip functions vanish beyond some cut-off. The horizontal dotted
line shows the exact ground-state energy $E_0$. The vertical dashed
lines mark the threshold values $a$ of blip grid spacing (see
text).}
\end{figure}

Up to now, we have considered only the rather artificial case where
the Fourier components of the blip function are strictly zero beyond
a cut-off. This means that the blip function must extend to infinity
in real space, and this is clearly no use if we wish to do all
calculations in real space. We actually want $f_0 (x)$ to be strictly
zero beyond some real-space cut-off $b_0$: $f_0 (x) = 0$ if
$| x | > b_0$. This means that $\hat{f} (q)$ will extend to
infinity in reciprocal space. However, we can expect that with a
judicious choice for the form of $f_0 (x)$ the Fourier components
$\hat{f} (q)$ will still fall off very rapidly, so that the
behavior of the total energy is still essentially as shown in Fig.~1.
If the choice is not judicious, we shall need a considerably
greater effort to bring $E$ within a specified tolerance of the
exact value than if we were using a plane-wave basis set. With
a plane-wave basis, a certain cut-off $G_{\rm max}$ is needed in
order to achieve a specified tolerance in $E$. With blip functions
whose Fourier components cut off at $G_{\rm max}$, we should need
a blip spacing of $\pi / G_{\rm max}$ to achieve the same tolerance. Our
requirement on the actual choice of blip function is that the spacing
needed to achieve the given tolerance should be not much less than
$\pi / G_{\rm max}$.

\subsection{B-splines as blip functions}
Given that the blip function cuts off in real space at some
distance $b_0$, it is helpful if the function and some of its
derivatives go smoothly to zero at this distance. If $f_0 (x)$
and all its derivatives up to and including the $n$th vanish
at $| x | = b_0$, then $\hat{f} (q)$ falls off asymptotically
as $1 / | q |^{n+2}$ as $| q | \rightarrow \infty$. One way of
making a given set of derivatives vanish is to build $f_0 (x)$
piecewise out of suitable polynomials. As an example, we examine here
the choice of $f_0 (x)$ as a B-spline.

B-splines are localized polynomial basis functions that are equivalent to
a representation of functions in terms of cubic splines. A single
B-spline $B (x)$ centered at the origin and covering the region
$| x | \leq 2$ is built out of third-degree polynomials in the four
intervals $-2 \leq x \leq -1$, $-1 \leq x \leq 0$, $0 \leq x \leq 1$
and $1 \leq x \leq 2$, and is defined as:
\begin{eqnarray}
B (x) =  \left\{
\begin{array}{ccr}
1 - \frac{3}{2} x^2 + \frac{3}{4} | x |^3 \; & \mbox{  if  } &
0 < | x | < 1  \\
       \frac{1}{4} ( 2 - | x | )^3 \; & \mbox{  if  } &
1 < | x | < 2 \\
      0 \; & \mbox{  if  } & 2 < | x |  
\end{array}
\right. 
\end{eqnarray}
The function and its first two derivatives are continuous everywhere.
The Fourier transform of $B (x)$, defined as in Eq.~(\ref{eq:fourier}),
is:
\begin{equation}
\hat{B} (q) = \frac{1}{q^4} ( 3 - 4 \cos q + \cos 2q ) \; ,
\end{equation}
which falls off asymptotically as $q^4$, as expected. Our choice of blip
function is thus $f_0 (x) = B (2x / b_0 )$, so that
$\hat{f} (q) = \hat{B} ( \frac{1}{2} b_0 q )$.

The transform $\hat{B} (q)$ falls rapidly to small values in the
region $| q | \simeq \pi$. It is exactly zero at the set of wavevectors
$q_n = 2 \pi n$ ($n$ is a non-zero integer), and is very small in a
rather broad region around each $q_n$, because the lowest
non-vanishing term in a polynomial expansion of $3 - 4 \cos q + \cos 2q$
is of degree $q^4$. This suggests that this choice of blip function
will behave rather similarly to one having a Fourier cut-off
$q_{\rm cut} = 2 \pi / b_0$. In other words, if we keep
$b_0$ fixed and reduce the blip spacing $a$, the energy should
approach the value obtained in a plane-wave calculation having
$G_{\rm max} = q_{\rm cut}$ when $a \simeq \frac{1}{2} b_0$. The
practical tests that now follow will confirm this. (We note that
B-splines are usually employed with a blip spacing equal to 
$\frac{1}{2} b_0$;
here, however, we are allowing the spacing to vary).

\subsection{Practical tests}
Up to now, it was convenient to work in one dimension, but for practical
tests we clearly want to go to real three-dimensional systems.
To do this, we simply take
the blip function $f_0 ( {\bf r} )$ to be the product of factors
depending on the three Cartesian components of ${\bf r}$: 
\begin{equation}
f_0 ( {\bf r} ) = p_0 ( x )\, p_0 ( y )\, p_0 ( z ) \; .
\label{eq:factor}
\end{equation}
All the considerations outlined above for a blip basis in one
dimension apply unchanged to the individual factors $p_0 ( x )$ etc,
which are taken here to be B-splines.
Corresponding to the blip grid of spacing $a$ in one dimension, the
blip functions $f_0 ( {\bf r} )$ now sit on the points of a
three-dimensional grid which we assume here to be simple cubic. The
statements made above about the properties of the B-spline
basis are expected to remain true in this three-dimensional form.

We present here some tests on the performance of a B-spline basis
for crystalline Si. At first
sight, it might appear necessary to write a new code in order
to perform such tests. However, it turns out that rather minor
modifications to an existing plane-wave code allow one to produce
results that are identical to those that would be obtained
with a B-spline basis. For the purpose of the present tests, this
is sufficient. The notion behind this device is that blip functions
can be expanded in terms of plane waves, so that the function-space
spanned by a blip basis is contained within the space spanned by a
plane-wave basis, provided the latter has a large enough $G_{\rm max}$.
Then all we have to do to get the blip basis is to project
from the large plane-wave space into the blip space. Mathematical
details of how to do this projection in practice are given in
the Appendix. The practical tests have been done with the 
CASTEP code~\cite{payne},
which we have modified to perform the necessary projections.

Our tests have been done on the diamond-structure Si crystal, using the
Appelbaum-Hamann~\cite{appelbaum:hamann}
 local pseudopotential; this is an empirical
pseudopotential, but suffices to illustrate the points of
principle at issue here.
The choice of
$k$-point sampling is not expected to make much difference
to the performance of blip functions, and we have done the calculations
with a $k$-point set corresponding to the lowest-order 4 $k$-point
Monkhorst-Pack~\cite{monkhorst:pack} sampling for an 8-atom cubic cell.
If we
go the next-order set of 32 $k$-points, the total energy per
atom changes by less than 0.1~eV. For reference purposes, we
have first used CASTEP in its normal unmodified plane-wave form
to examine the convergence of total energy with respect to plane-wave
cut-off for the Appelbaum-Hamann potential. We find that for
plane-wave cut-off energies $E_{\rm pw} = \hbar^2 G_{\rm max}^2 / 2 m$
equal to 150, 250 and 350~eV, the total energies per atom are --115.52,
--115.64 and --115.65~eV. This means that to obtain an accuracy of
0.1~eV/atom, a cut-off of 150~eV (corresponding to $G_{\rm max} =
6.31$~\AA$^{-1}$) is adequate. According to the discussion of Sec.~2.2,
the properties of this plane-wave basis should be quite well 
reproduced by a blip-function basis of B-splines having half-width
$b_0 = 2 \pi / G_{\rm max} = 1.0$~\AA, and
the total energy calculated with this basis should converge
rapidly to the plane-wave result when the blip
spacing falls below $a \simeq \frac{1}{2} b_0 = 0.5$~\AA.

\begin{figure}[tbh]
\begin{center}
\leavevmode
\epsfxsize=8cm
\epsfbox{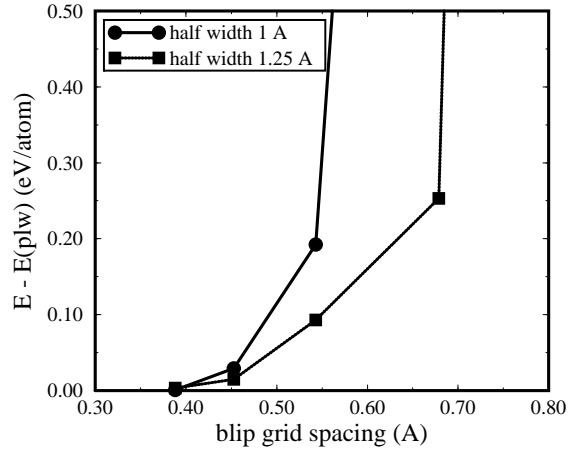}
\end{center}
\caption{Convergence of the total energy for two different blip
half-widths
as a function of the blip grid spacing. The calculations were
performed with
the Appelbaum-Hamann pseudopotential. $E(plw)$ is
the plane-wave result, obtained with a cutoff of 250~eV.}
\end{figure}

We have done the tests with two widths of B-splines: $b_0 =$
1.25 and 1.0~\AA, and in each case we have calculated $E$
as a function of the blip spacing $a$. In all cases, we have used a
plane-wave cut-off large enough to ensure that errors in
representing the blip-function basis are negligible compared
withe errors attributable to the basis itself. Our results for
$E$ as a function of $a$ (Fig.~2) fully confirm our expectations.
First, they have the general form indicated in Fig.~1. The difference
is that since we have no sharp cut-off in reciprocal space, $E$ does
not become constant when $a$ falls below a threshold, but instead continues to
decrease towards the exact value. Second, for $b_0 = 1.0$~\AA, $E$ does
indeed converge rapidly to the plane-wave result when $a$ falls
below $\sim 0.5$~\AA. Third, the larger blip width gives the lower
energy at larger spacings.

\section{The blip-function basis in linear-scaling calculations}
\subsection{Technical matters}
In our linear-scaling scheme, the support functions $\phi_\alpha ({\bf r})$
must be varied within support regions, which are centered on the
atoms. These regions, taken to be spherical with radius $R_{\rm reg}$
in our previous work, move with the atoms. 
In the present work, the $\phi_\alpha$
are represented in terms of blip functions. Each atom has a blip grid attached
to it, and this grid moves rigidly with the atom. The blip functions
sit on the point of this moving grid. To make the region localized,
the set of blip-grid points is restricted to those for which the
associated blip function is wholly contained within the region
radius $R_{\rm reg}$.
If we denote by $f_{\alpha \ell} ({\bf r})$
the $\ell$th blip function in the region supporting
$\phi_\alpha$, then the representation is:
\begin{equation}
\phi_\alpha ({\bf r}) = \sum_\ell b_{\alpha \ell} \,
f_{\alpha \ell} ({\bf r}) \; ,
\end{equation}
and the blip coefficients $b_{\alpha \ell}$ have to be varied
to minimize the total energy.

The $\phi_\alpha$ enter the calculation through their overlap matrix elements
and the matrix elements of kinetic and potential energies. The overlap
matrix $S_{\alpha \beta}$ [see Eq.~(\ref{eq:overlap})] can be expressed 
analytically
in terms of the blip coefficients:
\begin{equation}
S_{\alpha \beta} = \sum_{\ell \ell^\prime} b_{\alpha \ell} \,
b_{\beta \ell^\prime}\, s_{\alpha \ell , \beta \ell^\prime} \; ,
\end{equation}
where $s_{\alpha \ell , \beta \ell^\prime}$ is the overlap matrix between
blip functions:
\begin{equation}
s_{\alpha \ell , \beta \ell^\prime} = \int \! \! \mbox{d} {\bf r} \,
f_{\alpha \ell} \, f_{\beta \ell^\prime} \; ,
\end{equation}
which is known analytically. Similarly, the kinetic energy matrix elements:
\begin{equation}
T_{\alpha \beta} = - \frac{\hbar^2}{2 m} \int \! \! \mbox{d} {\bf r} \,
\phi_\alpha \nabla^2 \phi_\beta 
\end{equation}
can be calculated analytically by writing:
\begin{equation}
T_{\alpha \beta} = \sum_{\ell \ell^\prime} b_{\alpha \ell} \,
b_{\beta \ell^\prime} \, t_{\alpha \ell , \beta \ell^\prime} \; ,
\end{equation}
where:
\begin{equation}
t_{\alpha \ell , \beta \ell^\prime} = - \frac{\hbar^2}{2 m}
\int \! \! \mbox{d} {\bf r} \, f_{\alpha \ell} 
\nabla^2 f_{\beta \ell^\prime} \; .
\end{equation}

However, matrix elements of the potential energy cannot be
treated analytically, and their integrations must be approximated
by summation on a grid. This `integration grid' is, of course,
completely distinct from the blip grids. It does not move with
the atoms, but is a single grid fixed in space. If the position
of the $m$th point on the integration grid is called ${\bf r}_m$,
then the matrix elements of the local potential (pseudopotential plus
Hartree and exchange-correlation potentials) are approximated by:
\begin{equation}
V_{\alpha \beta} = \int \! \! \mbox{d} {\bf r} \, 
\phi_\alpha V \phi_\beta \simeq
\delta \omega_{\rm int} \sum_m \phi_\alpha ( {\bf r}_m )
V( {\bf r}_m ) \phi_\beta ( {\bf r}_m ) \; ,
\end{equation}
where $\delta \omega_{\rm int}$ is the volume per grid point. For a non-local
pseudopotential, we assume the real-space version of the 
Kleinman-Bylander~\cite{kleinman:bylander}
representation, and the terms in this are also calculated as a sum over
points of the integration grid. We note that the approximate equivalence
of the B-spline and plane-wave bases discussed above gives us an 
expectation for the required integration-grid spacing {\em h\/}. In a 
plane-wave calculation, {\em h\/} should in principle be less than
$\pi(2G_{\rm max})$. But the blip-grid spacing is approximately 
$a=\pi G_{\rm max}$. We therefore expect to need 
$h \approx \frac{1}{2} a$.

In order to calculate $V_{\alpha \beta}$ like this, we have to know all the
values $\phi_\alpha ( {\bf r}_m )$ on the integration grid:
\begin{equation}
\phi_\alpha ( {\bf r}_m ) = \sum_\ell b_{\alpha \ell}  \, 
f_{\alpha \ell} ( {\bf r}_m ) \; .
\label{eq:support}
\end{equation}
At first sight, it would seem that each point ${\bf r}_m$ would be
within range of a large number of blip functions, so that many
terms would have to be summed over for each ${\bf r}_m$
in Eq.~(\ref{eq:support}). In fact, this is
not so, provided the blip functions factorize into Cartesian components
in the way shown in Eq.~(\ref{eq:factor}).
To see this, assume that the blip grid and integration grid are cubic,
let the blip-grid index $\ell$ correspond to the triplet 
$( \ell_x , \ell_y , \ell_z )$, and let the factorization of 
$f_\ell ( {\bf r} )$ be written as:
\begin{equation}
f_\ell ( {\bf r}_m ) = p_{\ell_x} ( x_m ) \,  p_{\ell_y} ( y_m ) \,
p_{\ell_z} (z_m ) \; ,
\end{equation}
where $x_m$, $y_m$ and $z_m$ are the Cartesian components of ${\bf r}_m$
(we suppress the index $\alpha$ for brevity). The sum 
over $\ell$ in Eq.~(\ref{eq:support}) can then be performed as a 
sequence of three summations,
the first of which is:
\begin{equation}
\theta_{\ell_y \ell_z} ( x_m ) = \sum_{\ell_x} 
b_{\ell_x \ell_y \ell_z} \,  p_{\ell_x} ( x_m ) \; .
\end{equation}
The number of operations needed to calculate all these quantities
$\theta_{\ell_y \ell_z} ( x_m )$ is just the number of points
$( \ell_x , \ell_y , \ell_z )$ on the blip grid times the number 
$\nu_{\rm int}$ of points $x_m$ for which $p_{\ell_x} ( x_m )$ is
non-zero for a given $\ell_x$. This number $\nu_{\rm int}$ will generally
be rather moderate, but the crucial point is that the number of
operations involved is proportional only to $\nu_{\rm int}$ and not to
$\nu_{\rm int}^3$. Similar considerations will apply to the sums
over $\ell_y$ and $\ell_z$.

It is worth remarking that since we have to calculate
$\phi_\alpha ( {\bf r}_m )$ anyway, we have the option of calculating
$S_{\alpha \beta}$ by direct summation on the grid as well. In fact,
$T_{\alpha \beta}$ can also be treated this way, though here
one must be more careful, since it is essential that its symmetry
($T_{\alpha \beta} = T_{\beta \alpha}$) be preserved by whatever scheme
we use. This can be achieved by, for example, calculating the gradient
$\nabla \phi_\alpha ( {\bf r}_m )$ analytically on the integration
grid and then using integration by parts to express $T_{\alpha \beta}$
as an integral over $\nabla \phi_\alpha \cdot \nabla \phi_\beta$.
In the present work, we use the analytic forms for 
$s_{\alpha \ell , \beta \ell^{\prime}}$ and 
$t_{\alpha \ell , \beta \ell^{\prime}}$.

A full linear-scaling calculation requires minimization of the
total energy with respect to the quantities $\phi_{\alpha}$ and
$L_{\alpha \beta}$. However, at present we are concerned solely
with the representation of $\phi_{\alpha}$, and the cut-off 
applied to $L_{\alpha \beta}$ is irrelevant. For our practical
tests of the blip-function basis, we have therefore taken the
$L_{\alpha \beta}$ cut-off to infinity, which is equivalent to exact
diagonalization of the Kohn-Sham equation. Apart from this, the
procedure we use for determining the ground state, i.e. minimizing
$E$ with respect to the $\phi_{\alpha}$ functions, is essentially
the same as in our previous 
work~\cite{hernandez:gillan,hernandez:gillan:goringe}. We use 
conjugate-gradients~\cite{numerical:recipes} minimization
with respect to the blip-function coefficients $b_{\alpha \ell}$.
Expressions for the required derivatives are straightforward to derive
using the methods outlined earlier~\cite{hernandez:gillan:goringe}.

\subsection{Practical tests}
We present here numerical tests both for the 
Appelbaum-Hamann~\cite{appelbaum:hamann} local
pseudopotential for Si used in Sec.~2 and for a standard 
Kerker~\cite{kerker}
non-local pseudopotential for Si. The aims of the tests are: first,
to show that the B-spline basis gives the accuracy in support-function
calculations to be expected from our plane-wave calculations; and second
to examine the convergence of $E$ towards the exact plane-wave
results as the region radius $R_{\rm reg}$ is increased. For present
purposes, it is not particularly relevant to perform the tests on
large systems. The tests have been done on perfect-crystal Si at
the equilibrium lattice parameter, as in Sec.~2.3.

\begin{table}[tbh]
\begin{center}
\begin{tabular}{cccc}
& $h$ (\AA) & $E$ (eV/atom) & \\
\hline
& 0.4525 & -0.25565 & \\
& 0.3394 &  0.04880 & \\
& 0.2715 &  0.00818 & \\
& 0.2263 & -0.01485 & \\
& 0.1940 &  0.00002 & \\
& 0.1697 & -0.00270 & \\
& 0.1508 &  0.00000 & \\
\end{tabular}
\end{center}
\caption{Total energy $E$ as a function of integration grid spacing
$h$ using
a region radius $R_{\rm reg} = 2.715$~\AA. The blip half-width
$b_0$ was set to
0.905~\AA\ and the blip grid spacing used was 0.4525~\AA. The zero of
energy is set equal to the result obtained with the finest grid.}
\end{table}

We have shown in Sec.~2.3 that a basis of B-splines having a
half-width $b_0 = 1.0$~\AA\ gives an error of $\sim 0.1$~eV/atom
if the blip spacing is $\sim 0.45$~\AA. 
For the present tests we have used the
similar values $b_0 = 0.905$~\AA\ and $a = 0.4525$~\AA, in the
expectation of getting this level of agreement with CASTEP plane-wave
calculations in the limit $R_{\rm reg} \rightarrow \infty$.
To check the influence of integration grid spacing $h$, we have made
a set of
calculations at different $h$ using $R_{\rm reg} = 2.715$~\AA, which is
large enough to be representative (see below). The results (Table 1)
show that $E$ converges rapidly with decreasing $h$, and 
ceases to vary for present purposes when $h = 0.194$~\AA.
This confirms our expectation that $h \approx \frac{1}{2} a$.  We have then
used this grid spacing to study the variation of $E$ with $R_{\rm reg}$,
the results for which are given in Table~2, where we also
compare with the plane-wave results. The extremely rapid convergence
of $E$ when $R_{\rm reg}$ exceeds $\sim 3.2$~\AA\ is very striking,
and our results show that $R_{\rm reg}$ values yielding an accuracy
of $10^{-3}$~eV/atom are easily attainable. The close agreement with
the plane-wave result fully confirms the effectiveness of the blip-function
basis. As expected from the variational principle, $E$ from the blip-function
calculations in the $R_{\rm reg} \rightarrow \infty$ limit lies
slightly above the plane-wave value, and the discrepancy of $\sim 0.1$~eV
is of the size expected from the tests of Sec.~2.3 for (nearly) the
present blip-function width and spacing. (We also remark in
parenthesis that the absolute agreement between results obtained
with two entirely different codes is useful
evidence for the technical correctness of our codes.)

\begin{table}[tbh]
\begin{center}
\begin{tabular}{ccc}
$R_{\rm reg}$ (\AA) & \multicolumn{2}{c}{$E$ (eV/atom)} \\
& local pseudopotential & non-local psuedopotential \\
\hline
2.2625 & 1.8659 & 1.9653 \\
2.7150 & 0.1554 & 0.1507 \\
3.1675 & 0.0559 & 0.0396 \\
3.6200 & 0.0558 & 0.0396 \\
4.0725 & 0.0558 & 0.0396 \\
\end{tabular}
\end{center}
\caption{Convergence of the total energy $E$ as a function of the
region radius $R_{\rm reg}$ for silicon with a local and a non-local
pseudopotential. The calculations were performed with a blip grid
spacing of 0.4525~\AA\ and a blip half-width of 0.905~\AA\ in both
cases.
The zero of energy was taken to be the plane wave result obtained with
each pseudopotential, with plane wave cutoffs
of 250 and 200 eV respectively.}
\end{table}

The results obtained in our very similar tests using the
Kleinman-Bylander form of the Kerker pseudopotential for Si are also
shown in Table~2. In plane-wave calculations, the plane-wave cut-off
needed for the Kerker potential to obtain a given accuracy is very similar
to that needed in the Appelbaum-Hamann potential, and we have therefore
used the same B-spline parameters. Tests on the integration-grid spacing
show that we can use the value $h = 0.226$~\AA, which is close to what
we have used with the local pseudopotential.
The total
energy converges in the same rapid manner for $R_{\rm reg} > 3.2$~\AA,
and the agreement of the converged result with the CASTEP value
is also similar to what we saw with the Appelbaum-Hamann pseudopotential.

\section{Discussion}
In exploring the question of basis sets for linear-scaling
calculations, we have laid great stress on the relation with
plane-wave basis sets. One reason for doing this is that the plane-wave
technique is the canonical method for pseudopotential calculations,
and provides the easiest way of generating definitive results by
going to basis-set convergence. We have shown that within the linear-scaling
approach the total energy can be taken to convergence by systematically
reducing the width and spacing of a blip-function basis set, just as
it can be taken to convergence by increasing the plane-wave cut-off in
the canonical method. By analyzing the relation between the plane-wave and
blip-function bases, we have also given simple formulas for estimating
the blip width and spacing needed to achieve the same accuracy as
a given plane-wave cut-off. In addition, we have shown that the density
of integration-grid points relates to the number of blip functions
in the same way as it relates to the number of plane waves. Finally,
we have seen that the blip-function basis provides a practical way of
representing support functions in linear-scaling calculations, and that
the total energy converges to the plane-wave result as the region
radius is increased.

These results give useful insight into what can be expected of linear-scaling
DFT calculations. For large systems, the plane-wave method requires
a massive redundancy of information: it describes the space of occupied
states using a number of variables of order $N \times M$ ($N$ the number of
occupied orbitals, $M$ the number of plane waves), whereas the number of
variables in a linear-scaling method is only of order $N \times m$
($m$ the number of basis functions for each support function). This means
that the linear-scaling method needs fewer variables than the plane-wave
method by a factor $m / M$. But we have demonstrated that to achieve
a given accuracy the number of blip functions per unit volume is not
much greater than the number of plane waves per unit volume. 
Then the factor $m / M$ is
roughly the ratio between the volume of a support region and the
volume of the entire system. The support volume must clearly depend on
the nature of the system. But for the Si system, we have seen that
convergence is extremely rapid once the region radius
exceeds $\sim 3.2$~\AA, corresponding to a region volume of 137~\AA$^3$,
which is about 7 times greater than the volume per atom of 20~\AA$^3$.
In this example, then, the plane-wave method needs more variables
than the linear-scaling method when the number of atoms $N_{\rm atom}$
is greater than $\sim 7$, and for larger systems it needs more
variables by a `redundancy factor' of $\sim N_{\rm atom} / 7$. (For
a system of 700 atoms, e.g., the plane-wave redundancy factor would
be $\sim 100$.) In this sense, plane-wave calculations on large systems
are grossly inefficient. However, one should be aware that there are
other factors in the situation, like the number of iterations needed
to reach the ground state in the two methods. We are not yet in a position
to say anything useful about this, but we plan to return to it.

Finally, we note an interesting question. The impressive rate of convergence
of ground-state energy with increase of region radius  shown in Table~2
raises the question of what governs this convergence rate, and whether it
will be found in other systems, including metals. We remark that
this is not the same as the well-known question about the rate of
decay of the density matrix $\rho ( {\bf r} , {\bf r}^{\prime} )$
as $| {\bf r} - {\bf r}^{\prime} | \rightarrow \infty$, because in our
formulation both $\phi_\alpha ( {\bf r} )$ and $L_{\alpha \beta}$
play a role in the decay. Our intuition is that this decay is controlled
by $L_{\alpha \beta}$. We hope soon to report results on the support
functions for different materials, which will shed light on this.

\section*{Acknowledgements}
This project was performed in the framework of the U.K. Car-Parrinello
Consortium, and the work of CMG is funded by the 
High Performance Computing Initiative (HPCI) through grant GR/K41649. 
The work of EH is supported by EPSRC grant GR/J01967.
The use of B-splines
in the work arose from discussions with James Annett.

\appendix

\section*{Using plane-wave calculations to test blip functions}
We explain here in more detail how a standard plane-wave code can be used
to test the performance of blip-function basis sets (see Sec. 2.3).

In a plane-wave calculation, the occupied orbitals $\psi_i ({\bf r})$
are expressed as:
\begin{equation}
\psi_i ({\bf r}) = \Omega^{-1/2} \sum_{\bf G} c_{i {\bf G}}
\exp ( i {\bf G} \cdot {\bf r} ) \; ,
\end{equation}
where ${\bf G}$ goes over all reciprocal lattice vectors whose
magnitude is less than the plane-wave cut-off $G_{\rm max}$, and $\Omega$
is the volume per cell. (We assume $\Gamma$-point sampling for simplicity.)
The usual procedure is to minimize the total energy $E$ with
respect to the plane-wave coefficients $c_{i {\bf G}}$, subject to the
orthonormality constraints:
\begin{equation}
\sum_{\bf G} c_{i {\bf G}}^\ast c_{j {\bf G}} = \delta_{ij} \; .
\end{equation}

In a blip-function basis, the orbitals can be regarded as expanded
in terms of the blip waves $\chi_{\bf G} ( {\bf r} )$ defined by:
\begin{equation}
\chi_{\bf G} ( {\bf r} ) = A_{\bf G} \sum_\ell f_\ell ({\bf r})
\exp ( i {\bf G} \cdot {\bf R}_\ell ) \; ,
\end{equation}
where the sum goes over all blip-grid points in one repeating cell,
and {\bf G} is any of the reciprocal-lattice vectors of the
repeating cell in the Brillouin zone associated with the blip grid.
(The number of different ${\bf G}$ vectors labelling $\chi_{\bf G}$
is thus the same as the number of points of the blip grid in one cell.)
The $\chi_{\bf G} ({\bf r})$ are orthogonal for different
values of ${\bf G}$, and we assume that the constant $A_{\bf G}$
is chosen so that the $\chi_{\bf G} ({\bf r})$ are normalized:
\begin{equation}
\int \mbox{d} {\bf r} \, \chi_{{\bf G}_1}^\ast \chi_{{\bf G}_2} =
\delta_{{\bf G}_1 , {\bf G}_2} \; .
\end{equation}
Then the orbitals are represented in terms of the blip waves as:
\begin{equation}
\psi_i ( {\bf r} ) = \sum_{\bf G} b_{i {\bf G}} \chi_{\bf G} ({\bf r}) \; ,
\end{equation}
and the ground state is determined by minimizing $E$ with
respect to the blip coefficients $b_{i {\bf G}}$ subject to 
orthonormality:
\begin{equation}
\sum_{\bf G} b_{i {\bf G}}^\ast b_{j {\bf G}} = \delta_{i j} \; .
\end{equation}

Now the $c_{i {\bf G}}$ can be expressed in terms of the $b_{i {\bf G}}$
by using the expansion of $\chi_{\bf G} ({\bf r})$ in terms
of plane waves:
\begin{equation}
\chi_{\bf G} ({\bf r}) = A_{\bf G} \omega^{-1} \sum_{\bf \Gamma}
\hat{f} ( {\bf G} + {\bf \Gamma} ) 
\exp \left[ i ( {\bf G} + {\bf \Gamma} ) \cdot {\bf r} \right] \; ,
\end{equation}
where $\hat{f} ({\bf q})$ is the Fourier transform of $f_0 ({\bf r})$,
$\omega$ is the volume per blip-grid point,
and ${\bf \Gamma}$ goes over the reciprocal-lattice vectors of the
blip grid. From this, we have:
\begin{equation}
\psi_i ({\bf r}) = \omega^{-1} \sum_{\bf G} A_{\bf G}
b_{i {\bf G}} \sum_{\bf \Gamma} \hat{f} ( {\bf G} + {\bf \Gamma} )
\exp \left[ i ( {\bf G} + {\bf \Gamma} ) \cdot {\bf r} \right] \; ,
\end{equation}
so that:
\begin{equation}
c_{i \, {\bf G} + {\bf \Gamma}} = \Omega^{1/2} \omega^{-1}
A_{\bf G} b_{i {\bf G}} \hat{f} ( {\bf G} + {\bf \Gamma} ) \; .
\end{equation}
Note that in interpreting this equation, ${\bf G}$ must be taken in the
first Brillouin zone of the blip grid.

This equation implies a relation between the plane-wave coefficients
$c_{i \, {\bf G} + {\bf \Gamma}}$ for the same values of ${\bf G}$
but different ${\bf \Gamma}$. We can therefore reproduce the effect
of a blip-function basis set by performing a standard plane-wave
calculation in which the $c_{i \, {\bf G} + {\bf \Gamma}}$
are constrained to obey this relation. Note that in practice the plane-wave
calculation must involve a plane-wave cut-off, so that the 
$c_{i \, {\bf G} + {\bf \Gamma}}$ will actually be set to zero
when $| {\bf G} + {\bf \Gamma} |$ exceeds this cut-off. This is
equivalent to setting $\hat{f} ( {\bf G} + {\bf \Gamma} )$ equal to
zero beyond the cut-off, and this amounts to a small smoothing of the
blip function in real space. Allowing for this modification, the space
spanned by the blip functions is a sub-space of the space spanned by the
plane waves. The constraint on the $c_{i \, {\bf G} + {\bf \Gamma}}$
is therefore a linear relation between these coefficients which
expresses the fact that their component orthogonal to the sub-space
vanishes. It can be written as:
\begin{eqnarray}
0 = c_{i \, {\bf G} + {\bf \Gamma}} - \hat{f} ( {\bf G} + {\bf \Gamma} )
\sum_{{\bf \Gamma}^\prime} \hat{f}^\ast ( {\bf G} + {\bf \Gamma}^\prime )
c_{i \, {\bf G} + {\bf \Gamma}^\prime} \left/ \right. \nonumber \\
\sum_{{\bf \Gamma}^\prime}
| \hat{f} ( {\bf G} + {\bf \Gamma}^\prime ) |^2 \; .
\end{eqnarray}

The modifications to the plane-wave code are therefore very simple. First,
the initial guess for the orbitals must be chosen so as to satisfy
this condition. Second, when the $c_{i \, {\bf G} + {\bf \Gamma}}$
are updated on the iterative path to the self-consistent ground state,
the changes of $c_{i \, {\bf G} + {\bf \Gamma}}$ must
be projected so as to satisfy this condition. This means
that if the unconstrained calculation would have produced the changes
$\delta c_{i \, {\bf G} + {\bf \Gamma}}$, these must be replaced by
$\delta c_{i \, {\bf G} + {\bf \Gamma}}^\parallel$ given by:
\begin{eqnarray}
\delta c_{i \, {\bf G} + {\bf \Gamma}}^\parallel =
\hat{f} ( {\bf G} + {\bf \Gamma} ) 
\sum_{{\bf \Gamma}^\prime} \hat{f}^\ast ( {\bf G} + {\bf \Gamma}^\prime )
\delta c_{i \, {\bf G} + {\bf \Gamma}^\prime} \left/ \right. \nonumber \\
\sum_{{\bf \Gamma}^\prime}
\left| \hat{f} ( {\bf G} + {\bf \Gamma}^\prime ) \right|^2 \; .
\end{eqnarray}
This will produce exactly the same result as if we had worked directly
with blip functions.

\end{document}